\documentclass[aps,prl,twocolumn,showpacs]{revtex4}
\usepackage{amsmath}
\usepackage{epsfig}
\usepackage{amssymb}
\usepackage{dcolumn}
\usepackage{graphicx}
\usepackage{dcolumn}

\begin{document}
\date{\today}

\title{Matter-wave solitons supported by dissipation}

\author{Adrian Alexandrescu}
\affiliation{Departamento de Matem\'aticas, Escuela T\'ecnica
Superior de Ingenieros Industriales, 
Universidad de Castilla-La Mancha 13071 Ciudad Real, Spain}
 
\author{V\'{\i}ctor M. P\'erez-Garc\'{\i}a}
\email{victor.perezgarcia@uclm.es}
\affiliation{Departamento de Matem\'aticas, Escuela T\'ecnica
Superior de Ingenieros Industriales, 
Universidad de Castilla-La Mancha 13071 Ciudad Real, Spain}

\begin{abstract}
We show how novel types of long-lived self-localized matter waves can be constructed with Bose-Einstein condensates. The ingredients leading to such structures are a spatial
phase generating a flux of atoms towards the condensate center and the dissipative mechanism provided by the inelastic three-body collisions in atomic 
Bose-Einstein condensates. The outcome is an striking example of \emph{nonlinear structure supported by dissipation}.
\end{abstract}


\maketitle

The realization of Bose-Einstein condensation with ultracold atomic gases has opened the door for many spectacular 
realizations of matter waves. In particular, the elastic two-body collisions between condensed atoms provide an effective 
nonlinear interaction within the atomic cloud. These nonlinear interactions have been used to obtain  different types of
 self-localized matter waves \cite{dark} denoted as matter-wave solitons.
 
 Most proposed and all experimentally realized solitons in ultracold atomic gases lead to self-localization along one 
 spatial dimension while in the others they must be  externally confined by magnetic or optical means.
Truly multidimensional solitons cannot be supported by the nonlinear interactions arising from elastic two-body collisions 
and nonlinearity is not able to balance robustly dispersion these scenarios \cite{collapse,Sulem}. 

In Bose-Einstein condensates (BECs) with Feschbach resonance management it has been predicted that certain two-dimensional 
solitons may exist \cite{Ueda,pisaBoris,Gaspar} but the idea does not extend trivially to three spatial dimensions without the adition of 
extra (external) confining potentials.

Another physical effect coming from the interaction between atoms in the condensate is dissipation. Three-body collisions usually lead expulsion of
atoms from the condensate leading to an effective dissipation which is thought to be responsible for its finite lifetime.

Dissipation usually acts against self-localization since it tends to take the system closer to the linear situation where no stable nonlinear structures exist.
 However, in some contexts dissipation has been shown to play an stabilizing role. In Ref. \cite{Ueda2} the addition of a phenomenological Landau damping to the Feschbach-resonance managed model has been shown to enhance stability. Another example is nonspreading  (linear) wave packets with  external imaginary potential \cite{Ober}. 

In this letter we present novel nonlinear structures in Bose-Einstein condensates self-trapped by the effect of nonlinear \emph{dissipative terms}. The physical idea behind this paper is that imprinting  an appropriate spatially dependent phase on a BEC  leads to a flux of particles from its periphery to the center which compensates the particles lost by three-body inelastic collisions. The outcome is a 
 long-lived nonlinear structure supported by dissipation.
 
We will work in the mean field approximation in which a BEC is modelled by the Gross-Pitaevskii equation
\begin{equation}
i\hbar \frac{\partial\Psi}{\partial t} = -\frac{\hbar^2}{2m} \Delta \Psi + V\Psi + \frac{4\pi\hbar^2a_s}{m}|\Psi|^2 \Psi + i\Gamma_3 |\Psi|^4 \Psi
\label{gp_equation}
\end{equation}
where $V$ accounts for any external trapping potential, $a_s$ is the $s-$wave  scattering length for elastic two-body collisions 
and $\Gamma_3=-\hbar K_3/12$, $K_3$ being the thresholdless three-body recombination rate \cite{kohler}. Eq. (\ref{gp_equation}) has been used as a model to study collapse in BECs with attractive interactions \cite{cBEC1,cBEC2,cBEC3,cBEC4,cBEC5,cBEC6,cBEC3b}. However, the nonlinear structures to be studied in this paper are not related with collapse phenomena, a fact that we will stress by working out examples with  $a_s= 0$ or even $a_s>0$.
Models similar to Eq. (\ref{gp_equation})  arise in the propagation of optical beams in Kerr media with multiphoton absorption processes  \cite{Dubietis,Porras}.

First, we change to the dimensionless variables $\boldsymbol{r}\equiv \boldsymbol{x}/a_0$, $\tau \equiv t/T$ ($\nu_0=1/T$) and $\psi\equiv\sqrt{a_0^3}\Psi$, where 
$\nu_0=\hbar/a^2_0 m$, and $a_0$ is  a characteristic size of the BEC 
so that Eq. (\ref{gp_equation}) becomes (from now on we set $V=0$)
\begin{equation}
i\frac{\partial \psi}{\partial \tau}=-\frac{1}{2}\Delta\psi+ g|\psi|^2\psi+i\gamma|\psi|^4\psi,
\label{gp_dimensionless}
\end{equation} 
with $g= 4\pi a_s/a_0$ and $\gamma =-K_3/(12\nu_0 a_0^6)$.

To obtain stationary self-trapped solutions of Eq. (\ref{gp_dimensionless}) we first write it in the modulus-phase representation, i. e. $\psi(\boldsymbol{r},\tau)=A(\boldsymbol{r},\tau)\exp[i\phi(\boldsymbol{r},\tau)],$ with $A(\boldsymbol{r},\tau)>0$
\begin{subequations}
\label{mod-pha}
\begin{eqnarray}
\frac{\partial (A^2)}{\partial \tau} & = & -\nabla \cdot \left(A^2\nabla \phi\right) + 2 \gamma A^6, \\
\frac{\partial \phi}{\partial \tau} & = & \frac{1}{2} \left[\frac{1}{A}\Delta A - \left(\nabla \phi\right)^2\right] -g A^2.
\end{eqnarray}
\end{subequations}
Stationary solutions of Eq. (\ref{gp_dimensionless}) satisfy  $\partial_{\tau} A^2=0$ and from Eq. (\ref{mod-pha}b) 
we get $\phi(\boldsymbol{r},\tau)=\varphi(\boldsymbol{r})-\delta \tau$, with $\delta>0$. 
Rescaling the spatial variables with $\boldsymbol{\rho}\equiv\sqrt{2\delta}\boldsymbol{r}$ we find 
\begin{subequations}
\label{mod-pha-st_normalized}
\begin{eqnarray}
 -2 (\nabla_{\boldsymbol{\rho}} A) (\nabla_{\boldsymbol{\rho}} \phi) - A\Delta_{\boldsymbol{\rho}} \phi  +  \frac{\gamma}{\delta} A^5 =   0, \label{phase_equation}\\
 \Delta_{\boldsymbol{\rho}} A - \left(\nabla_{\boldsymbol{\rho}} \phi\right)^2A -\frac{g}{\delta} A^3 + A  =  0. \label{amplitude_equation}
\end{eqnarray}
\end{subequations} 
 In this paper we will focus on solutions with radially symmetric amplitude $A(\boldsymbol{\rho}) = R(\rho)$. In quasi-two dimensional 
situations we will study solutions with phase given by $\phi(\boldsymbol{\rho},t) = \Phi (\rho) + m\theta$, being $\theta =  \text{atan}(\rho_2/\rho_1)$ i.e. including a 
possible vorticity. In three spatial dimensions we will restrict our attention to spherically symmetric stationary solutions $\phi(\boldsymbol{\rho},t) = \Phi (\rho)$. Then, 
Eq. (\ref{phase_equation}) can be integrated by using the divergence theorem and we get 
\begin{equation}\label{Phi}
\Phi'(\rho) = \frac{\gamma}{R^2\rho^{d-1}\delta} \int_0^\rho r^{d-1}R(r)^6 dr.
\end{equation}
Eq. (\ref{Phi}) allows us to obtain the asymptotic  behavior of the phase, $\Phi'(\rho) \approx  \tfrac{\gamma}{2\delta} R^4(\rho) \rho,$ when  $\rho \ll 1$, and 
$\Phi'(\rho)  \approx  Q/\left[\rho^{d-1}R^2(\rho)\right]$, for $\rho \gg 1$, where  we assume $Q =\tfrac{\gamma}{\delta}  \int_0^{\infty} r^{d-1} R^6 dr$ to be a finite quantity. 
The amplitude equation \eqref{amplitude_equation} becomes
\begin{eqnarray}\label{r-phase}
R'' + \tfrac{d-1}{\rho}R' - \left[(\Phi')^2 + \tfrac{m^2}{\rho^2}\right]R - \frac{g}{\delta} R^3 + R  =  0, \label{r-R}
\end{eqnarray}
plus $R(\infty) = 0$ and $R(0) = R_0$ $(m=0)$ or $R'(\rho) =  0$ $(m\neq 0)$. In the later case, it is easy to obtain 
$R(\rho) \sim R'(0) \rho^{|m|}$ for small $\rho$.
For large $\rho$ we get
\begin{equation}\label{asymp}
R'' + \tfrac{d-1}{\rho} R' + q R \simeq 0,
\end{equation} 
which gives $R(\rho) \sim 1/\rho^{(d-1)/2}$ asymptotically.

To obtain the following order in the approximation of $R(\rho)$ for large $\rho$ we define $h(\rho) = R(\rho)\rho^{(d-1)/2}$, which satisfies the Ermakov-Pinney equation  
\cite{Pinney}
 $h'' + h = Q^2/h^3$ whose solution can be found in explicit form
\begin{equation}\label{Ras}
R(\rho) \sim \frac{\sqrt{\frac{Q^2}{C_1^2} \cos^2 \rho + \left( C_1 \cos \rho + C_2 \sin \rho\right)^2}}{\rho^{(d-1)/2}}, \ \rho \gg 1.
\end{equation}
When $d=2$ this result is similar to that of Ref. \cite{Porras} with the different definition of $Q$.
Eq. (\ref{Ras}) guarantees the finiteness of $Q$ which was assumed previously.

\begin{figure}
\epsfig{file=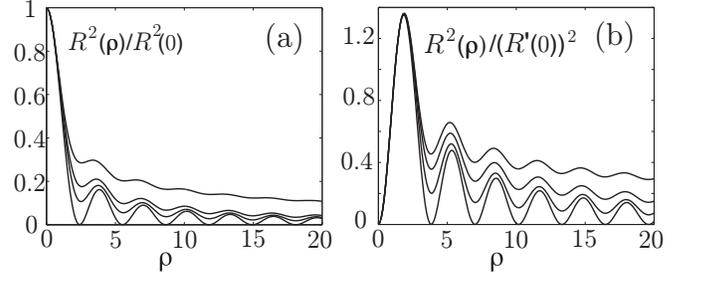,width=\columnwidth}
\caption{Radial profile of stationary solutions of Eq. (\ref{r-phase}) with $g=0$ and 
(a) $\alpha = \delta/\left[\gamma R^4(0)\right]=0.55,0.65,0.8,\infty$ (from higher to lower ones). 
(b) Stationary radial profile of vortex solutions with $m=\pm 1$ for $\delta/\gamma [R^\prime (0)]^4=2.9, 3.2, 4.1, \infty$.}
\label{prima}
\end{figure}

\begin{figure}
\epsfig{file=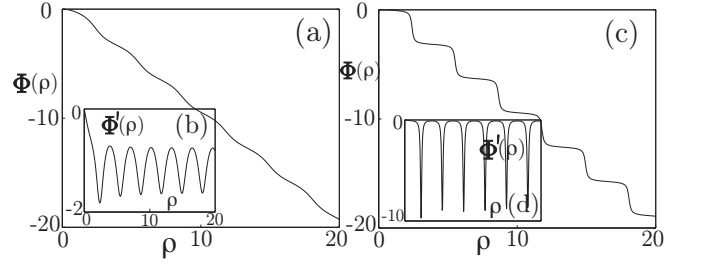,width=\columnwidth}
\caption{Phase and its derivative for solutions of Eq. (\ref{r-phase}) with $g=0$ and 
(a-b) $\alpha = \delta/\left[\gamma R^4(0)\right]=0.65$ (c-d) $\alpha = 5$.}
\label{phase}
\end{figure}

\begin{figure}
\epsfig{file=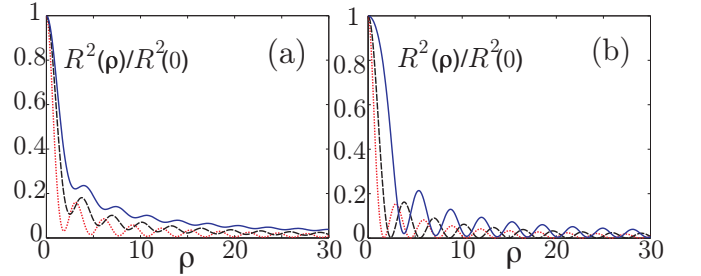,width=\columnwidth}
\caption{[Color online] Radial profile of solutions of Eq. (\ref{r-phase}) with (a) $\delta/\left[\gamma R^4(0)\right]=0.8$ and 
$gR^2(0)/\delta = -1.3$ (dotted line), 0 (dashed line), and 0.35 (solid line). (b) $\delta/\left[\gamma R^4(0)\right]=5$, $gR^2(0)/\delta = -2$ (dotted line), 0 (dashed line), and 0.85 (solid line).}
\label{second}
\end{figure}

\begin{figure}
\epsfig{file=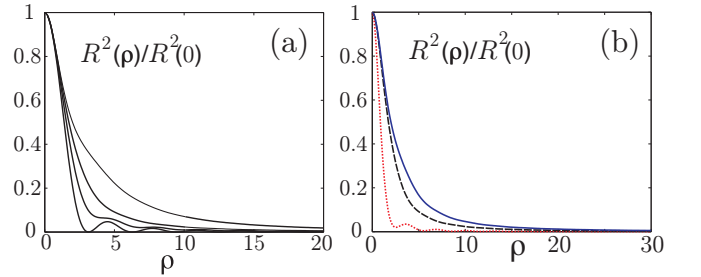,width=\columnwidth}
\caption{[Color online] Radial profile of solutions of Eq. (\ref{r-phase}) for $d=3$ with (a) $\delta/\left[\gamma R^4(0)\right]=0.36,0.4,0.5,\infty$ and 
$g=0$ (b) $\delta/\left[\gamma R^4(0)\right]=0.4$ and $gR^2(0)/\delta = -2$ (dotted line), 0 (dashed line), and 0.1 (solid line).}
\label{tercer}
\end{figure}

 We have solved numerically Eqs. \eqref{phase_equation} and (\ref{r-phase}) to find specific stationary configurations. Some results for $g = 0$ are shown in 
Fig. \ref{prima} for $m=0$ [Figs. \ref{prima}(a) and \ref{phase}] and $m=1$ [Fig. \ref{prima}(b)] as a function of the shooting parameter $ \alpha =  \delta/\gamma R^4(0)$. 
Solutions are found above a certain threshold value  $\alpha > \alpha_{th} = 0.54,\: 2.93,\: 0.36$ corresponding to the 2D solution without and with vorticity ($m=1$), and 3D solution, respectively.
When the effect of the nonlinearity is included (see Fig. \ref{second}) we obtain a compression of the solutions for $g<0$  [Fig. \ref{second}(a)]  and an expansion leading to slower decay for $g>0$  [Fig. \ref{second}(b)]. The threshold $\alpha_{th}$  is lowered when $g<0$and increased when $g>0$ there being a maximum
positive value $g_* = \delta/R^2(0)$ above which stationary solutions do not exist. 
It is remarkable that \emph{stationary solutions supported by dissipation can exist even when 
the scattering length is positive.}
In three spatial dimensions  we find a similar behavior with a faster decay of the amplitude (Fig. \ref{tercer}). 
In all cases the oscillatory behavior of the solutions and their decay are well described by Eq. (\ref{Ras})

 \emph{How can stationary solutions exist in a system with dissipation?} From the previous asymptotic analysis 
$R(\rho) \sim 1/\rho^{(d-1)/2}$, thus we can estimate the number of particles as $N \sim \int_0^{\infty} r^{d-1}R^2 dr = \infty$.  Moreover
\begin{equation}\label{losa}
\frac{d}{dt}N = 2\gamma \int_{\mathbb{R}^d} |\psi|^6 d\boldsymbol{r} < \infty.
\end{equation} 
Intuitively speaking, it is the fact that the solution has an infinite number of particles together with a \emph{finite} particle loss given by the right hand side of Eq. 
(\ref{losa}) which allows for the existence of stationary solutions even in the presence of three-body recombination losses.
\begin{figure}
\epsfig{file=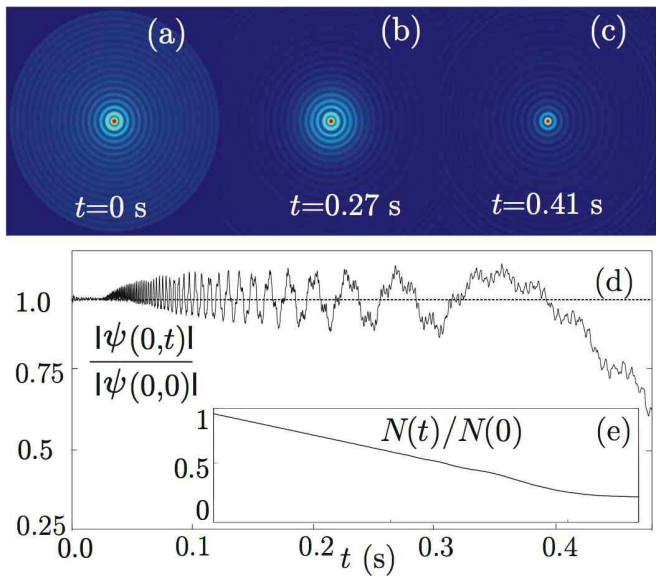,width=\columnwidth}
\caption{[Color online] Time evolution of a stationary solution of Eq. (\ref{gp_dimensionless}) for 
$\gamma = -2\times 10^{-10}$ (from $K_3=2\times10^{-26}$cm$^6$/s for $^7$Li), $g=0$, 
$\delta = 0.008$, $R_0 = 77$. The solution is solved in a square region of 1.2 mm size and 
the stationary solution is set to zero outside a disk of radius  450 $\mu$m which leaves about
$N_0=5.7\times 10^7$ atoms inside. (a-c) Pseudocolor plot showing $|\psi(x,y,t)|$ for different times. 
(d-e) Evolution of the (d) amplitude of the wavefunction and (e) number of particles in the condensate.
\label{evolution}}
\end{figure}

However, realistic solutions must have a finite number of particles. A simple way  to construct realistic quasi-stationary solutions is to cut the 
stationary solutions at an specific point far enough from the origin. In Fig. \ref{evolution} we show simulations of the evolution of one of those modified stationary 
solutions. The central soliton survives for very long times with quasi-stationary amplitude as shown in Fig. \ref{evolution}(d). This is a consequence of the phase structure of stationary solutions (see Fig. \ref{phase}), which 
leads to a particle flux towards the condensate center and thus induce a refilling mechanism of the central soliton which is manifest in Fig. \ref{evolution} (a-c). The rings surrounding the central peak play the role of a reservoir of atoms and constantly feed the central peak and dissapear progressively as time proceeds. From the practical point of view and since the amplitude of the rings is small what one would observe is a very long lived soliton lasting for times of the order of the condensate lifetime. 

\begin{figure}
\epsfig{file=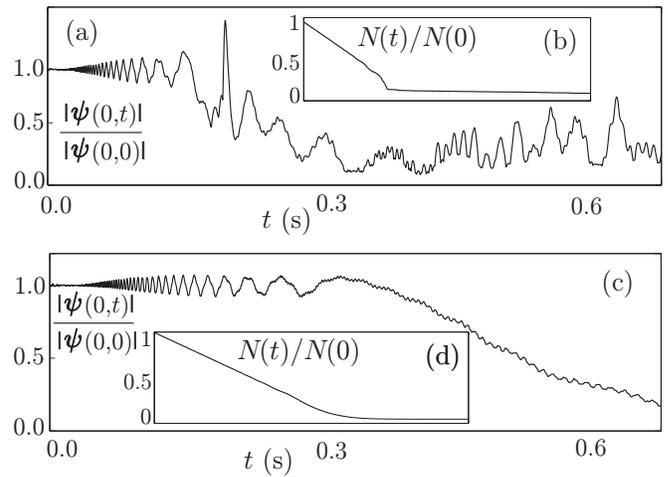,width=\columnwidth}
\caption{[Color online] Evolution of a stationary solution of Eq. (\ref{gp_dimensionless}) computed with
$\delta=10^{-2}$, $\gamma = -2\times 10^{-10}$.
Shown are the evolutions of the amplitude (a,c) and number of particles (b,d) in the condensate for (a-b) $g = -5.7\times 10^{-6}$, $R_0=132$,  on a square of side 800 $\mu m$, with the solution cut at 300 $\mu m$, leading to $N_0=7.6\times 10^7$ atoms
and (c-d) $g = 7.4\times10^{-7}$, $R_0=82$,  on a square physical domain of side 1.2 mm 
 with the profile set to zero outside a disk of  radius 450 $\mu$m which implies $N_0=1.2\times 10^8$ atoms.
\label{segunda-top}}
\end{figure}

The number of atoms used in our simulations are about an order of magnitude above those present in realistic experiments, however what it is really important is the shooting parameter $\delta/\left[\gamma R^4(0)\right]$. By choosing smaller values for $R(0)$ one may get smaller particle numbers but at the price of decreasing $\delta$ thus changing the solution and widening the spatial scales. These are options which must be taken on the basis of particular experimental scenarios.

\begin{figure}
\epsfig{file=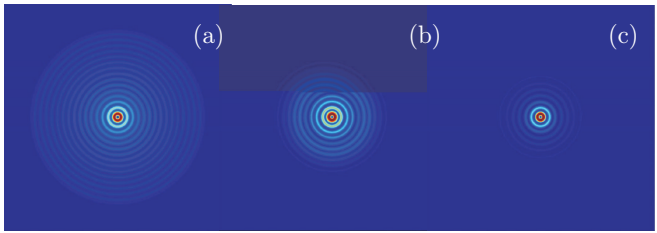,width=\columnwidth}
\caption{[Color online]  (a-c) Pseudocolor plots showing the evolution of $|\psi(x,y,t)|^2$ for different time values of a vortex type,$m=1$, solution computed with 
$g=0$, $\delta=10^{-2}$, $R_{max}=57$, on a square of side 900 $ \mu$m, and setting the profile to zero outside a disk of radius 350 $\mu$m. The total number of atoms is $N_0=7.6\times 10^7$.
\label{vortice}}
\end{figure}

We have checked that the same mechanism works for vortex type solutions as shown in Fig. \ref{vortice}.

The scenario described above is also valid for nonzero (positive and negative) values of $g$ as shown in  Fig. \ref{segunda-top}. This means that the phenomenon described here is not a result of  elastic two-body collisions. Moreover it seems that the phenomenon results is a very  long-lived central soliton even in condensates with positive scattering length, which is a very surprising result. 

We have not found any signs of the azimuthal 
instabilities predicted in Ref. \cite{Porras} for  systems with higher order dissipation. This fact requires further investigacion and could be either  because of a suppression of this instability in our situation  or just a consequence of a finite lifetime of our structure, smaller than the typical time for the instability to set in.

\emph{Experimental generation.-} To generate these structures in real experiments it would be very important to obtain appropriate initial data with amplitude and, more 
importantly, phase close to those corresponding to stationary solutions. In quasi-2D condensates phase imprinting methods \cite{Ph1,Ph2,Ph3,Ph4} can be used. In fact, imprinting  radially symmetric phases is simpler to do using absorption plates than originally proposed for vortices \cite{Ph1}.  As to the amplitude profile it could be achieved by using a Bessel  beam instead of a gaussian beam for trapping the condensate before releasing it.

In fully three-dimensional Bose-Einstein condensates phase imprinting methods would be more difficult to apply. However, three dimensional systems with attractive scattering 
length and three-body recombination in the collapsing regime develop the phenomenon known as superstrong collapse \cite{Sulem,Zakharov,dissipation}  which has been ignored in the physical literature studying collapse in Bose-Einstein condensates. This means that  a collapsing condensate spontaneously develops a structure of shells in 
both the amplitude and  phase similar to those shown in Fig. \ref{prima}  constantly feeding atoms to the region of higher density of atoms. In fact, those structures are present  in numerical simulations of collapse in Bose-Einstein condensates \cite{cBEC1,cBEC3,cBEC3b}. 
The phenomenon of spontaneous phase self-modulation in superstrong collapse provides a way for generating the matter-wave solitons presented in this paper. If during the initial stage of a 
blow-up event  the scattering length is changed to a subcritical value the violent compression and high losses associated to collapse would be suppressed but the superimposed phase structure could evolve to one of the attractors of the system, i.e. the stable stationary states.
This procedure could be a way to generate the matter wave soliton supported by dissipation proposed in this paper.

This work has been supported by grants BFM2003-02832 (Ministerio de Educaci\'on y Ciencia, MEC)
and PAI-05-001 (Consejer\'{\i}a de Educaci\'on y Ciencia de la Junta de Comunidades de Castilla-La Mancha).
A.A. acknowledges support from MEC through the scholarship AP-2004-7043.

\end{document}